\documentclass[aps,floatfix,groupedaddress,superscriptaddress,prl,twocolumn]{revtex4-1}
\usepackage[english]{babel}
\usepackage{graphicx}
\usepackage{amsmath,bbm,amssymb,multirow,times,bm,braket,mathtools,appendix}

\usepackage{color}
\definecolor{Ad}{rgb}{0.,0.3,0.7}
\definecolor{Pr}{rgb}{0.4,0.3,0.9}
\definecolor{Out}{rgb}{0.2,0.6,0.3}

\newcommand{\eu}{\mathrm{e}}
\newcommand{\im}{\mathrm{i}\,}

\begin{document}

\title{Quantum singular value decomposition of non-sparse low-rank matrices}
\author{Patrick Rebentrost}
\email{rebentr@mit.edu}
\affiliation{Research Laboratory of Electronics,
Massachusetts Institute of Technology, Cambridge, MA 02139}
\author{Adrian Steffens}
\affiliation{Dahlem Center for Complex Quantum Systems, Freie Universit\"at Berlin, 14195 Berlin}
\affiliation{Research Laboratory of Electronics,
Massachusetts Institute of Technology, Cambridge, MA 02139}
\author{Seth Lloyd}
\email{slloyd@mit.edu}
\affiliation{Research Laboratory of Electronics,
Massachusetts Institute of Technology, Cambridge, MA 02139}
\affiliation{Department of Mechanical Engineering, Massachusetts Institute of Technology, Cambridge, MA 02139}

\begin{abstract}
In this work, we present a method to exponentiate non-sparse indefinite low-rank 
matrices on a quantum computer.  
Given an operation for accessing the elements of the matrix, our method allows 
 singular values and associated singular vectors to be found quantum mechanically 
 in a time exponentially 
faster in the dimension of the matrix than known classical algorithms. 
The method extends to non-Hermitian and non-square matrices via embedding matrices.
In the context of the generic singular value decomposition of a matrix, we discuss 
the Procrustes problem of finding a closest isometry to a given matrix.
\end{abstract}

\maketitle

Matrix computations are central to many algorithms in optimization and machine learning \cite{Golub2012,Boyd2004,Murphy2012}. At the heart of these algorithms regularly lies an eigenvalue or a singular value decomposition of a matrix, or a matrix inversion. Such tasks could be performed efficiently via phase estimation on a universal quantum computer \cite{Nielsen2010}, as long as the matrix can be simulated (exponentiated) efficiently and controllably as a Hamiltonian acting on a quantum state. Almost exactly twenty years ago, Ref.~\cite{Lloyd1996} paved the way for such a simulation of quantum systems by introducing an efficient algorithm for exponentiating Hamiltonians with tensor product structure---enabling applications such as in quantum computing for quantum chemistry 
\cite{Aspuru-Guzik2005}. Step by step, more general types of quantum systems were tackled and performance increased: Aharonov and Ta-Shma~\cite{Aharonov2003} showed a method for simulating quantum systems described by sparse Hamiltonians, while Childs {\it et al.}~\cite{Childs2003} demonstrated the simulation of a quantum walk on a sparse graph. Berry  {\it et al.}~\cite{Berry2007} reduced the temporal scaling to approximately linear via higher-order Suzuki integrators. Further improvements in the sparsity scaling were presented in Ref.~\cite{Berry2012}.
Beyond sparse Hamiltonians, quantum principal component analysis (qPCA) was shown to handle non-sparse positive semidefinite low-rank Hamiltonians~\cite{Lloyd2014} when given multiple copies of the Hamiltonian as a quantum density matrix.
This method has applications in quantum process tomography and state discrimination~\cite{Lloyd2014}, as well as in quantum machine learning~\cite{Wiebe2014,Wiebe2014_2,Benedetti2015,Schuld2015,Schuld2016,Kerenidis2016,Lau2016},  specifically in curve fitting~\cite{Wang2014} and support vector machines \cite{Rebentrost2014}. 
In an oracular setting, Ref.~\cite{Childs2010,Childs2011,Berry2012} showed the simulation of non-sparse Hamiltonians via discrete quantum walks. 
The scaling in terms of the simulated time $t$ is $t^{3/2}$ or even linear in $t$.

In the spirit of  Ref.~\cite{Lloyd2014}, we provide an alternative method for non-sparse matrices in an oracular setting which requires only one-sparse simulation techniques. We achieve a run time in terms of the matrix maximum element and a $t^2$ scaling. We discuss a class of matrices with low-rank properties that make the non-sparse methods
efficient. 
Compared to Ref.~\cite{Lloyd2014} the matrices need not be positive semidefinite. 
In order to effectively treat a general non-Hermitian non-quadratic matrix, we 
make use of an indefinite ``extended Hermitian matrix'' that incorporates the 
original matrix. With such an extended matrix, we are able to efficiently determine the singular value decomposition of dense non-square, low-rank matrices. As one possible application of our method, we discuss 
the Procrustes problem~\cite{Golub2012} of finding a closest isometric matrix.

\paragraph{Method.}
We have been given an $N \times N$ dense (non-sparse)
Hermitian indefinite matrix $A\in\mathbb{C}^{N\times N}$ via efficient oracle access to the elements of $A$. The oracle either performs an efficient computation of the matrix elements or provides access to a storage medium for the elements such as quantum RAM \cite{Giovannetti2008,Giovannetti2008_2}. 
Our new method simulates $\eu^{-\im (A/N) t}$ on an arbitrary quantum state for  arbitrary times $t$. Note that the eigenvalues of $A/N$ are bounded by $\pm  \Vert A \Vert_{\max} $, where   $\Vert A \Vert_{\max}$ is the maximal absolute value of the matrix elements of $A$. 
This means that there exist matrices $A$ for which the unitary $\eu^{-\im (A/N) t}$ can be far from the identity operator for a time of order $\Vert A \Vert_{\max}^{-1}$, 
i.e.~an initial state can evolve to a perfectly distinguishable state.  
For such times, the unitary $\eu^{-\im (A/N) t}$ can be well approximated by a unitary generated by a low-rank matrix. 

Let $\sigma$ and $\rho$ be $N$-dimensional density matrices.
The state $\sigma$ is the target state on which the matrix exponential of $A/N$ is applied to, while multiple copies of  $\rho$ are used as ancillary states.
Our method embeds the $N^2$ elements of $A$ into a 
Hermitian sparse matrix $S_A\in\mathbb{C}^{N^2 \times N^2}$, which we call ``modified swap matrix''
because of its close relation to the usual swap matrix. Each column of $S_A$ contains a single element of $A$. 
The modified swap matrix between the registers for a single copy of $\rho$ and $\sigma$ is 
\begin{eqnarray}
S_A&=&\sum_{ j, k=1}^N A_{jk} |k\rangle \langle j| \otimes | j \rangle \langle k | \in\mathbb{C}^{N^2 \times N^2}.
\end{eqnarray}
This matrix is one-sparse in a quadratically bigger space and reduces to the usual swap matrix for $A_{jk}=1$ and $j,k=1,\dots,N$. 
Given efficient oracle access to the elements, we can simulate a one-sparse matrix such as $S_A$ 
with a constant number of oracle calls and negligible error \cite{Aharonov2003,Childs2003,Berry2007,Harrow2009}.
We discuss the oracle access below.
This matrix exponential of $S_A$ is applied to a tensor product of a uniform superposition and an arbitrary state. 
Performing $S_A$ for small $\Delta t$ leads to a reduced dynamics of
$\sigma$ when expanded to terms of second order in $\Delta t$ as
\begin{eqnarray}\label{eqMainExpansion}
{\rm tr}_1 \{ \eu^{- \im S_A \Delta t}\, \rho \otimes \sigma\, \eu^{ \im S_A \Delta t} \}&=& \\ \nonumber
\sigma -\im  {\rm tr}_1 \{ S_A\, \rho \otimes \sigma \} \Delta t  &+&\im {\rm tr}_1 \{ \rho \otimes \sigma \,S_A \}\Delta t  +O(\Delta t^2).
\end{eqnarray}
Here, ${\rm tr}_1$ denotes the partial trace over the first register containing $\rho$. 
The first $O(\Delta t)$ term is ${\rm tr}_1 \{ S_A\, \rho \otimes \sigma \} 
= \sum_{j,k=1}^N A_{jk} \langle j|\rho |k\rangle   | j \rangle \langle k |  \sigma. $
Choosing  $\rho= |\vec 1\rangle \langle \vec 1|$, with $|\vec 1\rangle\coloneqq \frac{1}{\sqrt{N}} \sum_k |k\rangle $
  the uniform superposition, leads to
${\rm tr}_1 \{ S_A\ \rho \otimes \sigma \} = \frac{A}{N}\, \sigma$.
This choice for $\rho$ contrasts with qPCA, where $\rho$ is proportional to the simulated matrix \cite{Lloyd2014}. Analogously, the second $O(\Delta t)$ term becomes
${\rm tr}_1 \{ \rho \otimes \sigma \ S_A \}= \sigma\,  \frac{A}{N}$.
Thus for small times, evolving with the modified swap matrix $S_A$ on the bigger system is equivalent to evolving with $A/N$  on the $\sigma$ subsystem,
\begin{eqnarray}\nonumber
{\rm tr}_1 \{ \eu^{- \im S_A \Delta t}\, \rho \otimes \sigma\, \eu^{ \im S_A \Delta t} \}&=&
\sigma -\im  \frac{\Delta t}{N}\, [A, \sigma]  +O(\Delta t^2) \\ &\approx& \eu^{-\im \frac{A}{N} \Delta t}\, \sigma\, \eu^{\im \frac{A}{N} \Delta t}.\label{eq:equiv}
\end{eqnarray}
Let  $\epsilon_0$ be the trace norm of the error term $O(\Delta t^2)$. We can bound this error by
$\epsilon_0 \leq 2 \Vert A \Vert_{\max}^2 \Delta t^2$ (see Appendix).
Here,  $\Vert A \Vert_{\max}=\max_{mn} |A_{mn}|$ denotes the maximal absolute element of $A$.
Note that $\Vert A \Vert_{\max}$ coincides with the largest absolute eigenvalue of $S_A$. 
The operation in Eq.~(\ref{eq:equiv}) can be performed multiple times in a forward Euler fashion using multiple copies of $\rho$. For $n$ steps  the resulting error is $\epsilon = n\, \epsilon_0$. The simulated time is $t=n\,\Delta t$. Hence, fixing $\epsilon$ and $t$,   
\begin{equation}\label{eqNumberOfSteps}
n= O\left (\frac{t^2  }{\epsilon} \Vert A \Vert_{\max}^2\right )
\end{equation}
steps are required to simulate $\eu^{-i \frac{A}{N} t}$. 
The total run time of our method is 
$n T_A$,
the number steps $n$ is multiplied with the matrix oracle access time $T_A$ (see below). 

We discuss for which matrices the algorithm runs efficiently.
Note that an upper bound  for the eigenvalues of $A/N$ in terms of the maximal matrix element is $|\lambda_j|/N \leq \Vert A \Vert_{\max}$.
At a simulation time $t$ only the eigenvalues of $A/N$ with $\vert \lambda_j\vert/N = \Omega(1/t)$ matter. Let the number of these eigenvalues be $r$.
Thus, effectively a matrix $A_r/N$ is simulated with 
${\rm tr} \{ A_r^2/N^2 \}=\sum_{j=1}^r \lambda_j^2/N^2= \Omega( r / t^2)$. It also holds that ${\rm tr} \{ A_r^2/N^2 \} \leq \Vert A \Vert_{\max} ^2$. Thus, the rank of the effectively simulated matrix is $r = O( \Vert A \Vert_{\max}^2 t^2)$.

Concretely, 
for the algorithm to be efficient in terms of matrix oracle calls, we require that the number of simulation steps $n$ is $O({\rm poly} \log N)$. Let the desired error be $1/\epsilon = O({\rm poly} \log N)$. 
Assuming $\Vert A \Vert_{\max}= \Theta(1)$, meaning a constant independent of $N$, we have from Eq.~(\ref{eqNumberOfSteps}) that we can only exponentiate for a time $t=O({\rm poly} \log N)$.
For such times, only the large eigenvalues of $A/N$ with $\vert \lambda_j\vert/N = \Omega(1/{\rm poly} \log N)$ matter.
Such eigenvalues can be achieved when the matrix is dense enough, 
for example $A/N$ has $\Theta(N)$ non-zeros of size $\Theta(1/N)$ per row.
For the rank of the simulated matrix in this case we find that $r =O({\rm poly} \log N)$, thus effectively a low-rank matrix is simulated.
To summarize, we expect the method to work well for low rank matrices $A$ that are dense with relatively small matrix elements. 

A large class of matrices satisfies these criteria. 
Sample a random unitary $U\in \mathbb{C}^{N\times N}$ and $r$ suitable eigenvalues of size $\vert \lambda_j \vert = \Theta(N)$ and multiply them as 
$U\,{\rm diag}_r (\lambda_j)\, U^\dagger$ to construct $A$.
Here, ${\rm diag}_r (\lambda_j)$ is the diagonal matrix with the $r$ eigenvalues on the diagonal and zero otherwise.
A typical random normalized vector has absolute matrix elements of size 
$O(1/\sqrt{N})$. The outer product of such a vector with itself has absolute matrix elements of size $O(1/N)$. 
Each eigenvalue of absolute size $\Theta(N)$ is multiplied with such an outer product
 and the $r$ terms are summed up. 
Thus, a typical matrix element of $A$ will be of size $O(\sqrt{r})$ and  $ \Vert A \Vert_{\max}=O(r)$. 

\paragraph{Phase estimation.}
Phase estimation provides a gateway from unitary simulation to many interesting applications. 
For the use in phase estimation, 
we extend our method such that the matrix exponentiation of $A/N$ can be performed conditioned on additional control qubits. With our method, the eigenvalues $\lambda_j/N$ of $A/N$ can be both positive and negative.
The modified swap operator $S_A$ for a Hermitian matrix $A$ with eigendecomposition $A=\sum_j \lambda_j |u_j\rangle \langle u_j|$ is augmented as $ \ket{1}\bra{1} \otimes S_A $, which
still is a one-sparse Hermitian operator.  The resulting unitary $\eu^{-\im \ket{1}\bra{1}\otimes S_A \Delta t}=\ket{0}\bra{0} \otimes \mathbbm{1} + 
\ket{1}\bra{1} \otimes \eu^{-\im S_A \Delta t}$ is efficiently simulatable.
This operator is applied to a state $|c\rangle \langle c|\otimes \rho \otimes \sigma $ where
$|c\rangle$ is an arbitrary control qubit state.
Sequential application of such controlled operations allows the use phase estimation to
prepare the state \cite{Harrow2009} 
\begin{equation} \label{eqPhaseEstimation}
|\phi\rangle = \frac{1}{\sqrt{\sum_j |\beta_j|^2}} \sum_{\frac{|\lambda_j|}{N} \geq \epsilon} \beta_j |u_j\rangle |\frac{\lambda_j}{N}\rangle
\end{equation} 
from an initial state $|\psi\rangle|0\hdots 0\rangle$ with $O(\lceil \log(1/ \epsilon) \rceil)$ control qubits forming an eigenvalue value register. 
Here, $\beta_j = \langle u_j| \psi \rangle$ 
and  $\epsilon$ is the accuracy for resolving eigenvalues. To achieve this accuracy,
phase estimation is run for a total time $t=O(1/\epsilon)$. 
Thus, $O(  \Vert A \Vert_{\max}^2 / \epsilon^3 )$ queries of the oracle for $A$ are required, which 
is of order $O({\rm poly} \log N)$ under the low-rank assumption for $A$ discussed above.

\paragraph{Matrix oracle and resource requirements.}
To simulate the modified swap matrix, we employ the methods developed in Refs.~\cite{Childs2003, Berry2007}. 
First, we assume access to the original matrix $A$, 
\begin{equation}\label{eqAOracle}
|j\,k\rangle |0\cdots 0\rangle \mapsto |j\,k\rangle |A_{jk}\rangle.
\end{equation}
This operation can be provided by quantum random access memory (qRAM) \cite{Giovannetti2008,Giovannetti2008_2} using $O(N^2)$ 
storage space and quantum switches for accessing the data in $T_{\rm A}=O(\log^2 N)$ operations. Alternatively, there matrices whose elements are efficiently computable, i.e.~  $T_{\rm A}= O({\rm poly} \log N)$. 
For the one-sparse matrix $S_A$, the unitary operation for the sparse simulation methods
\cite{Childs2003, Berry2007} can be simply constructed from the oracle in Eq.~(\ref{eqAOracle}) and
is given by~
\begin{equation}\label{eqSwapOracle}
|(j,k)\rangle |0\cdots 0\rangle \mapsto |(j,k)\rangle |(k,j), (S_A)_{(k,j),(j,k)}\rangle.
\end{equation}
Here, we use $(j,k)$ as label for the column/row index of the modified swap matrix. 

We compare the required resources with those of other methods for sparse and non-sparse matrices.
For a general $N\times N$ and $s$-sparse matrix, $O(sN)$ elements need to be stored. In certain cases, the 
sparse matrix features more structure and its elements can be computed efficiently \cite{Berry2007,Harrow2009}.
For non-sparse matrices and the qPCA method in Ref.~\cite{Lloyd2014}, only 
multiple copies of the density matrix as opposed to an
operation as in Eq.~(\ref{eqAOracle}) are required for applications such as state tomography. 
For machine learning via qPCA \cite{Lloyd2014,Rebentrost2014}, the
density matrix is prepared from a classical source via quantum RAM~\cite{Giovannetti2008,Giovannetti2008_2} and requires $O(N^2)$ storage. 
In comparison, the requirements of the method in this work are in principle not higher than these sparse and non-sparse methods, both in the case of qRAM access and in the case when matrix elements are computed instead of stored. 

\paragraph{Non-square matrices.}
Our method enables us also to determine properties of general non-square low-rank matrices effectively.
To determine the singular value decomposition of a matrix $A=U\Sigma V^{\dagger}\in\mathbb{C}^{M\times N}$ with rank $r$, simulating the positive semidefinite matrices $AA^{\dagger}$
and $A^{\dagger}A$ via qPCA yields the correct singular values and vectors. However, essential information is missing, leading to ambiguities in the singular vectors that become evident when inserting diagonal matrices into the singular value decomposition of $AA^{\dagger}$ that change the relative phases of the singular vectors,
\begin{eqnarray}
AA^{\dagger}
=U\Sigma^{2}U^{\dagger} 
=U \Sigma D^\dagger V^{\dagger}\ \,VD\Sigma U^{\dagger}
=:\hat{A}\hat{A}^{\dagger},
\end{eqnarray}
with $D\coloneqq\mathrm{diag}(\mathrm{e}^{-\mathrm{i}\vartheta_{j}})$, $\vartheta_{j}$ being arbitrary phases. If
$Av_{j}=\sigma_{j}u_{j}$ for each $j=1,\dots,r$, then
\begin{equation}
\hat{A}v_{j}=U\Sigma D^{\dagger}V^{\dagger}v_{j}=\sigma_{j}\mathrm{e}^{\mathrm{i}\vartheta_{j}}u_{j}\coloneqq\sigma_{j}\hat{u}_{j},
\end{equation}
which means different phase relations between left and right singular
vectors in $\hat{A}$ from those in $A$.
Although $A$ and $\hat{A}$ still share the same singular values
and even the same singular vectors up to phase factors, $\Vert A-\hat{A}\Vert_F$
will in general (with the exception of positive semidefinite matrices,
where $U=V$) not be zero or even be small: The matrix $A$ cannot
be reproduced this way---a singular value decomposition is more than a set of singular values
and normalized singular vectors. This affects all kinds of algorithms
that 
require the appropriate phase relations between each left
singular vector $u_{j}$ and the according right singular vector $v_{j}$.
Such applications are determining the best low-rank approximation
of a matrix, signal processing algorithms discussed in Ref.~\cite{Steffens2016}, or 
determining the nearest isometric matrix, related to the unitary Procrustes problem, of a non-Hermitian matrix.

In order to overcome this issue, consider the ``extended matrix" 
\begin{equation}\label{eqExtendedMatrix}
\tilde{A}\coloneqq\left[\begin{array}{cc}
0 & A\\
A^{\dagger} & 0
\end{array}\right],
\end{equation}
which was introduced for singular value computations in Ref.~\cite{Golub1965} and recently in sparse quantum matrix inversion in 
\cite{Harrow2009}.
The eigenvalues of $\tilde{A}$ correspond to $\{\pm\sigma_{j}\}$ with 
$\{\sigma_{j}\}$ being the singular values  of $A$ for $j=1,\dots,r$. 
The corresponding eigenvectors
are proportional to $(u_{j},\pm v_{j})\in\mathbb{C}^{M+N}$, see Appendix.
The left and right singular vectors of $A$ can be extracted
from the first $M$ and last $N$ entries, respectively. Since $\tilde{A}$
is Hermitian, its eigenvectors can assumed to be orthonormal:
$\Vert (u_{j},v_{j})\Vert ^{2}  =  \Vert u_{j}\Vert^{2}+\Vert v_{j}\Vert^{2}=1,$ and $(u_{j},v_{j})\cdot(u_{j},-v_{j})^{\dagger}=  \Vert u_{j}\Vert^{2}-\Vert v_{j}\Vert^{2}=0,$
from which follows that the norm of each of the subvectors $u_{j}$
and $v_{j}$ is $1/\sqrt{2}$, independent of their respective
lengths $M$ and $N$.
The important point is that the eigenvectors of the extended matrix
preserve the correct phase relations between the left and right singular
vectors since
$(\mathrm{e}^{\mathrm{i}\vartheta_{j}} u_{j}, v_{j})$  is
 only an eigenvector of $\tilde A$ for the correct phase $\mathrm{e}^{\mathrm{i}\vartheta_{j}}=1$. 

The requirements for our quantum algorithm can be satisfied also for the extended matrix. 
For randomly sampled left and right singular vectors,  the matrix elements have maximal size of $O(\sum_{j=1}^r \sigma_j/\sqrt{MN})$, thus $\sigma_j=O(\sqrt{MN})$.
In addition, an $1/(M+N)$ factor arises in the simulation of the extended matrix from the ancillary state 
$\rho=|\vec 1\rangle \langle \vec 1|$ as before, which leads to the requirement  $\sigma_j=\Theta(M+N)$.
These two conditions for $\sigma_j$ can be satisfied if the matrix $A$ is not too skewed, i.e. $M=\Theta(N)$.
In summary, by simulating the corresponding Hermitian extended matrices,
general complex matrices of low rank can be simulated efficiently,
yielding the correct singular value decomposition.

\paragraph{Procrustes problem.}

The unitary Procrustes problem is to find the unitary matrix that most accurately transforms one matrix into another. 
It has many applications, such as in shape/factor/image analysis and statistics~\cite{Golub2012}. 
We consider non-square matrices
thus consider the Procrustes problem to find the \textit{isometry} that most accurately transforms one matrix into another.
Formally, minimize $\Vert WB-C\Vert_F$ among all isometries $W\in\mathbb{C}^{M\times N}$, $W^\dagger W = \mathbbm{1}$, with $B\in\mathbb{C}^{N\times K}$ and $C\in\mathbb{C}^{M\times K}$, where  $M>N$.
The problem is equivalent to the general problem of finding the nearest isometric matrix 
$W\in\mathbb{C}^{M\times N}$ to a matrix $A\in\mathbb{C}^{M\times N}$ by taking $A= C B^\dagger$.
Since our quantum algorithm is restricted to low rank matrices, let $A=C B^\dagger$ be low-rank with rank $r$ and singular value decomposition $A=U\,\Sigma\, V^\dagger$ with $U\in\mathbb{C}^{M\times r}$, $\Sigma\in\mathbb{R}^{r\times r}$, and $V\in\mathbb{C}^{N\times r}$. 
The optimal solution to the Procrustes problem is $W=U\, V^\dagger$ \cite{Golub2012}, setting all singular values to one, in both the low-rank and the full-rank situation.
Since $A$ is assumed to be low rank, 
we find a \textit{partial} isometry with $W^\dagger W = \mathbbm{P}_{{\rm col} (V)}$, 
with $\mathbbm{P}_{{\rm col} (V)}$ the projector into the subspace spanned by 
the columns of $V$.
Thus, $W$ acts as an isometry for vectors in that subspace (see Appendix).

In a quantum algorithm, we want to apply the nearest low-rank isometry  to a quantum state 
$|\psi\rangle$.  The state $|\psi\rangle$ is assumed to be in or close to the subspace spanned by the columns of $V$. 
We assume that the extended matrix for $A$ in Eq.~(\ref{eqExtendedMatrix}) is given in oracular form
and that $A$ is not too skewed such that $\sigma_j/(M+N)=\Theta(1)$ and $\Vert A \Vert_{\max}=\Theta(1)$.
We perform phase estimation on the input state $|0,\psi\rangle|0\hdots0\rangle$ and, analogous to 
Eq.~(\ref{eqPhaseEstimation}), obtain a state proportional to 
\begin{equation}
 \smashoperator{\sum_{\frac{\sigma_j}{M+N} \geq  \epsilon}}~ \beta_j^\pm |u_j, \pm v_j\rangle | \pm \frac{\sigma_j}{M+N}\rangle
\end{equation}
with $\beta_j^\pm = \langle u_j,\pm v_j| 0,\psi \rangle = \pm \langle v_j| \psi \rangle/\sqrt{2}$. The sum has $2r$ terms corresponding to the eigenvalues of the extended matrix with absolute value greater than $(M+N) \epsilon$. 
Performing a $\sigma_z$ operation on the qubit encoding the sign of the respective eigenvalue an uncomputing the eigenvalue register yields a state proportional to 
$\sum_j~ \beta_j |u_j, \pm v_j\rangle$.
Projecting onto the $u_j$ part (success probability $1/2$) results in a state proportional to 
\begin{equation}
\smashoperator{\sum_{\frac{\sigma_j}{M+N} \geq \epsilon}}~ |u_j\rangle  \langle v_j| \psi \rangle \propto U\,V^\dagger |\psi\rangle.
\end{equation}
This prepares the desired state for  the non-square low-rank Procrustes problem with accuracy $\epsilon$ in runtime $O( \Vert A\Vert_{\max}^2 \log^2 (N+M)/\epsilon^3)$.
Classically, performing the singular value decomposition of a low-rank $A$ without further structural assumptions takes generally $O(N^3)$.

\paragraph{Conclusion.}
The method presented here 
allows  non-sparse low-rank non-positive Hermitian $N\times N$ matrices 
$A/N$ to be exponentiated for a time $t$ with accuracy $\epsilon$ in run time
$O\left( \frac{t^2}{\epsilon} \Vert A \Vert_{\max}^2 \,T_A\right)$, where $\Vert A \Vert_{\max} $ is the maximal absolute element of
$A$. The data access time is $T_A$. If the matrix elements are accessed via quantum RAM or computed efficiently and the significant eigenvalues of $A$ are $\Theta(N)$, our method can achieve a  run time of $O\left({ \rm poly} \log N \right )$ for a large class of matrices.
Our method allows non-Hermitian and non-square matrices to be exponentiated 
via extended Hermitian matrices. 

We have shown how compute the singular value decomposition 
of a non-Hermitian non-sparse matrix on a 
quantum computer directly while keeping all the correct relative phase information. 
As one of the many potential applications of the singular value decomposition, 
we can find the pseudoinverse of a matrix and the closest isometry exponentially faster 
than any known classical algorithm. 
It remains to be seen if the time complexity of our method can be improved from $O(t^2)$ to an approximately linear scaling via higher-order Suzuki-Trotter steps or other techniques. In addition, by using a (possibly unknown) ancillary state other than the uniform superposition, the oracular setting of the present work and the tomography setting of \cite{Lloyd2014} 
could be combined. 

\begin{acknowledgments}
We are grateful to Iman Marvian for insightful discussions.
We acknowledge support from DARPA, NSF, and AFOSR. 
AS thanks the German National Academic Foundation 
(Studienstiftung des deutschen Volkes) 
and the Fritz Haber Institute of the Max Planck Society for support.
\end{acknowledgments}

\bibliography{QML}


\appendix
 
\section{Appendix}

\paragraph {Norms.}
Denote the maximum absolute element of a matrix $A\in \mathbb{C}^{N\times N}$ with
$
\Vert A \Vert_{\max}:= \max_{j,k} |A_{jk}|.
$
The Frobenius or Hilbert-Schmidt norm is given by 
$
\Vert A \Vert_{F}:= \sqrt{\sum_{j,k} |A_{jk}|^2 }
$
and its nuclear norm by
$
\Vert A \Vert_\ast := \sum_{i=1}^r \sigma_i,
$
where $r$ is the rank and $\sigma_j$ are the singular values.

\paragraph{Modified swap matrix.}
The modified swap matrix is defined as
\begin{equation}
S_A=\sum_{ j, k=1}^N A_{jk} |k\rangle \langle j| \otimes | j \rangle \langle k | \in\mathbb{C}^{N^2 \times N^2}.
\end{equation}
Taking $A_{jk}\to 1$ leads to the original swap matrix 
$
S=\sum_{ j, k=1}^N |k\rangle \langle j| \otimes | j \rangle \langle k | \in\mathbb{C}^{N^2 \times N^2}.
$
The $N^2$ eigenvalues of $S_A$ are 
\begin{equation}
A_{11}, A_{22},\hdots,A_{NN}, A_{12},-A_{12},\hdots,A_{j,k>j},-A_{j,k>j},\hdots,
\end{equation}
where $k>j$ denotes an index $k$ greater than $j$.
The maximal absolute eigenvalue of $S_A$ is thus $\max_{j,k} |A_{jk}|\equiv \Vert A \Vert_{\max}$, corresponding to the maximal absolute matrix element of $A$.
The square of the modified swap matrix is
\begin{eqnarray}\label{eqSwapSquare}
(S_A)^2&=&\sum_{ j, k=1}^N |A_{jk}|^2\, |k\rangle \langle k| \otimes | j \rangle \langle j | \leq \Vert A \Vert_{\max}^2 \,\mathbbm{1}.
\end{eqnarray}
Its eigenvalues are $|A_{jk}|^2$ and the maximal eigenvalue is $\Vert A \Vert_{\max}^2$.
This already points to the result that the second order error of our method naturally scales with  $\Vert A \Vert_{\max}^2$, which we will now derive.

\paragraph{Error analysis.}
In the following, we  estimate the error from the second-order term in $\Delta t$ in the expansion Eq.~(\ref{eqMainExpansion}).
The nuclear norm of the operator part of the second order error is
\begin{eqnarray}\label{eq:error1}
\epsilon_{\rho,\sigma} = \Vert {\rm tr}_1 \{ S_A\ \rho \otimes \sigma \ S_A \} &-&\frac{1}{2} {\rm tr}_1 \{ (S_A)^2\ \rho \otimes \sigma \} \\ &-& \frac{1}{2}{\rm tr}_1 \{  \rho \otimes \sigma\ (S_A)^2 \} \Vert_\ast. \nonumber
\end{eqnarray}
In Ref.~\cite{Lloyd2014}, this error was equal to 
$\epsilon_{\rho,\sigma}^{\rm qPCA} = \Vert \rho - \sigma \Vert_\ast \leq 2$, 
which is achieved in the present algorithm by choosing $A$ such that $A_{jk}=1$ for each $j,k$. 
Here, our algorithm coincides with the qPCA method for $\rho$ chosen as the uniform superposition.
For general low-rank $A$, we bound Eq.~\eqref{eq:error1} via the triangle inequality.
Taking the nuclear norm of the first term results in
\begin{align}
\Vert {\rm tr}_1 \{ S_A \rho \otimes \sigma S_A \} \Vert_\ast \leq &  \Vert   S_A \rho \otimes \sigma S_A \Vert_\ast \nonumber \\
\leq&\Vert  \rho \otimes \sigma  \Vert_\ast \Vert S_A^2 \Vert_\ast \leq  \Vert A\Vert_{\max}^2.
\end{align}
The second and third term can be treated similarly. We obtain $\Vert {\rm tr}_1 \{ (S_A)^2 \rho \otimes \sigma \} \Vert_\ast \leq  \Vert A\Vert_{\max}^2$.
Combining all terms yields the bound
\begin{equation}
\epsilon_{\rho,\sigma} \leq 2\Vert A \Vert_{\max}^2.
\end{equation}

\vspace{.7cm}
\paragraph{Extended matrices.}
We define the Hermitian extended matrix $\tilde{A}$ of a complex-valued, not necessarily square matrix $A\in\mathbb{C}^{M\times N}$ as
\begin{equation}\label{}
\tilde{A}=\left[\begin{array}{cc}
0 & A\\
A^{\dagger} & 0
\end{array}\right]\in\mathbb{C}^{(M+N)\,\times\,(M+N)}.
\end{equation}
Using block matrix identities for the determinant, we obtain its characteristic
polynomial
\begin{equation}
\chi_{\tilde{A}}(\lambda)=\lambda^{\vert M-N\vert}\,\det\,(\lambda\mathbbm{1}+\sqrt{AA^{\dagger}})(\lambda\mathbbm{1}-\sqrt{AA^{\dagger}}).
\end{equation}
The eigenvalues of $\tilde{A}$ are either zero or correspond to $\{\pm\sigma_{j}\}$, the singular values of $A$ for $j=1,\dots,r$ with an additional sign. Hence, if $A$ has low rank $r$, then $\tilde{A}$ has low rank $2r$.
The corresponding eigenvectors
are proportional to $(u_{j},\pm v_{j})\in\mathbb{C}^{M+N}$ since
\begin{eqnarray}
\left[\begin{array}{cc}
\mp\sigma_{j}\mathbbm{1} & A\\
A^{\dagger} & \mp\sigma_{j}\mathbbm{1}
\end{array}\right]\cdot\left[\begin{array}{c}
u_{j}\\
\pm v_{j}
\end{array}\right]=0,
\end{eqnarray}
where $u_j$ and $v_j$ are the $j$th left and right singular vector of~$A$, respectively.
The important point is that the eigenvectors of the extended matrix
preserve the correct phase relations between the left and right singular
vectors since
$(\mathrm{e}^{\mathrm{i}\vartheta_{j}} u_{j}, \pm v_{j})$  is
 only an eigenvector of $\tilde A$ for the correct phase $\mathrm{e}^{\mathrm{i}\vartheta_{j}}=1$,
\begin{align}\nonumber
\left[\begin{array}{cc}
\mp\sigma_{j}\mathbbm{1} & A\\
A^{\dagger} & \mp\sigma_{j}\mathbbm{1}
\end{array}\right]
\!\cdot\!\left[\begin{array}{c}
\mathrm{e}^{\mathrm{i}\vartheta_{j}}u_{j}\\
\pm v_{j}
\end{array}\right]=&\left[\begin{array}{c}
\mp\sigma_{j}\mathrm{e}^{\mathrm{i}\vartheta_{j}}u_{j}\pm Av_{j}\\
\mathrm{e}^{\mathrm{i}\vartheta_{j}}A^{\dagger}u_{j}-\sigma_{j}v_{j}
\end{array}\right]\\=&(\mathrm{e}^{\mathrm{i}\vartheta_{j}}-1)\sigma_{j}\left[\begin{array}{c}
\mp u_{j}\\
 v_{j}
\end{array}\right].
\end{align}
The right hand side is only equal to zero for the correct phase $\mathrm{e}^{\mathrm{i}\vartheta_{j}}=1$. 

\paragraph{Low-rank Procrustes.}
Let the isometry be $W=U V^\dagger$ with $U\in\mathbb{C}^{M\times r}$ and $V\in\mathbb{C}^{N\times r}$.
Assume that $M>N$, giving orthogonal columns in the full-rank Procrustes problem ($r=N$).
We find for the low-rank (partial) isometry that
\begin{equation}
W^\dagger W = V U^\dagger U V^\dagger =V V^\dagger = \sum_{j=1}^r \vec v_j \vec v_j^\dagger.
\end{equation}
Pick an arbitrary vector 
$\vec x = \sum_{j=1}^r \alpha_j \vec v_j + \vec x^\perp= \vec x^\parallel+ \vec x^\perp$.
where $\vec x^\perp$ denotes the part orthogonal to the orthonormal vectors $\vec v_j$.
Then,
\begin{equation}
W^\dagger W \vec x = \sum_{j=1}^r \alpha_j \vec v_j = \vec x^\parallel.
\end{equation}
Thus, $W^\dagger W$ acts as the identity operator in the low-rank subspace, and projects out the space perpendicular to that subspace.

\end{document}